\documentclass[11pt]{article}

\usepackage{amsmath, amsfonts, mathrsfs,latexsym,color,array}

\usepackage{cite}
\usepackage{url}
\usepackage{ntheorem}

\definecolor{bluem}{rgb}{0,0,0.5}

\definecolor{mycolor}{cmyk}{0.5,0.1,0.5,0}
\definecolor{michel}{rgb}{0.5,0.9,0.9}

\definecolor{turquoise}{rgb}{0.25,0.8,0.7}
\definecolor{bluem}{rgb}{0,0,0.5}

\definecolor{MDB}{rgb}{0,0.08,0.45}
\definecolor{MyDarkBlue}{rgb}{0,0.08,0.45}

\definecolor{MLM}{cmyk}{0.1,0.8,0,0.1}
\definecolor{MyLightMagenta}{cmyk}{0.1,0.8,0,0.1}

\definecolor{HP}{rgb}{1,0.09,0.58}

{\catcode `\@=11 \global\let\AddToReset=\@addtoreset}
\AddToReset{equation}{section}

\newtheorem{Theorem} {\sc  Theorem\rm} [section]

\newtheorem{Lemma} [Theorem] {\sc  Lemma\rm}

\theorembodyfont{\upshape}
\newtheorem{Remark}[Theorem]{\sc Remark\rm}

\newcommand{\beqar}{\begin{deqarr}}
\newcommand{\eeqar}{\end{deqarr}}

\newcommand{\beaa}{\begin{eqnarray*}}
\newcommand{\eeaa}{\end{eqnarray*}}

\newcommand{\bel}[1]{\begin{equation}\label{#1}}
\newcommand{\bea}{\begin{eqnarray}}
\newcommand{\bean}{\begin{eqnarray}\nonumber}
\newcommand{\beal}[1]{\begin{eqnarray}\label{#1}}
\newcommand{\eea}{\end{eqnarray}}
\newcommand{\eeal}[1]{\label{#1}\end{eqnarray}}

\newcommand{\qed}{\hfill $\Box$ \medskip}
\newcommand{\proof}{\noindent {\sc Proof:\ }}
\newcommand{\be}{\begin{equation}}
\newcommand{\eeq}{\end{equation}}
\newcommand{\ee}{\end{equation}}
\newcommand{\beqa}{\begin{eqnarray}}
\newcommand{\eeqa}{\end{eqnarray}}
\newcommand{\beqan}{\begin{eqnarray*}}
\newcommand{\eeqan}{\end{eqnarray*}}
\newcommand{\ba}{\begin{array}}
\newcommand{\ea}{\end{array}}

\DeclareFontFamily{OT1}{rsfs}{} \DeclareFontShape{OT1}{rsfs}{m}{n}{
<-7> rsfs5 <7-10> rsfs7 <10-> rsfs10}{}
\DeclareMathAlphabet{\mycal}{OT1}{rsfs}{m}{n}

\newcommand{\bit}{\begin{itemize}}
\newcommand{\eit}{\end{itemize}}

\newcounter{shownewstuffflag}

\newcommand{\startnewstuff}{\ifnum\value{shownewstuffflag}>0\color{blue}\fi}
\newcommand{\finishnewstuff}{\ifnum\value{shownewstuffflag}>0\color{black}\fi}

\newcounter{oldeq}

\newcounter{mnotecount}[section]

\newcommand{\rmnote}[1]{}

\def\beq{\begin{equation}}
\def\eeq{\;. \end{equation}}

\newcommand{\eq}[1]{(\ref{#1})}

\begin{document}
\title{Singular Yamabe metrics and initial data with \emph{exactly} Kottler--Schwarzschild--de Sitter ends II. \\ Generic metrics}
\author{Piotr T. Chru\'sciel
\\  LMPT, F\'ed\'eration Denis Poisson, Tours \\ Mathematical Institute and Hertford College, Oxford
\\
\\
Frank Pacard
\\ Universit\'e Paris Est and Institut Universitaire de France
\\
\\
Daniel Pollack
\\ University of Washington
}


\maketitle

\abstract{We present a gluing construction which adds, via a localized deformation, \emph{exactly Delaunay ends} to \emph{generic} metrics with constant positive scalar curvature. This  provides time-symmetric initial data sets for the vacuum Einstein equations with positive cosmological constant with  \emph{exactly} Kottler--Schwarzschild--de Sitter ends, extending the results in~\cite{ChPollack}.}

\section{Introduction} \label{Sintro}

There exists very strong evidence suggesting that we live in a world with strictly positive cosmological constant $\Lambda$~\cite{WoodVasey:2007jb,Riess:2006fw}. This leads to a need for a better understanding of the space of solutions of Einstein equations with $\Lambda>0$. In~\cite{ChPollack} it has been shown how time-symmetric initial data for such space-times, which contain \emph{asymptotically Delaunay ends}, can be deformed to initial data with \emph{exactly Delaunay ends}. The resulting space-times contain  regions of infinite extent on which  the metric takes \emph{exactly} the Kottler--Schwarzschild--de Sitter form. Moreover, such ends can be used for creating wormholes, or connecting initial data sets, provided the neck parameters of the ends match.

The object of this short note is to show  that \emph{generic}  constant positive scalar curvature  metrics can be deformed, by a local deformation, to constant positive scalar curvature metrics containing asymptotically Delaunay ends.  The method of \cite{ChPollack} is then used to obtain \emph{exactly Delaunay ends}.

We further point out that the neck parameter of each new asymptotically Delaunay end can be arbitrarily prescribed within an interval
$(0,\epsilon_0)$, for some $\epsilon_0>0$. This flexibility in prescribing the neck sizes  guarantees that the exactly Delaunay ends  can be matched to perform gluings.

\section{Statement of the result}
\label{STt}

We begin with some terminology.  A \emph{static KID}  is a function $N$ satisfying
\bel{normKID}
D_iD_j N - N R_{ij} -\Delta_g N g_{ij}=0
\;,
\ee
We shall say that \emph{there exist no local static KIDs near $p$} if there exist sequences $0<\eta_i < \delta_i\to_{i\to \infty} 0$ such that there exist no nontrivial solution of \eq{normKID} on the annuli
\[
A_p(\eta_i, \delta_i): = \bar B_p(\delta_i) \setminus B_p(\eta_i) \, ,
\]
where $B_p(r)$ is a geodesic ball centered at $p$ of radius $r$. Observe that \eq{normKID} forms a set of overdetermined equations, it is natural to expect that for generic metrics on $M$ there  will not be any local static KIDs at any point $p\in M$. Precise statements to this effect have been proved in~\cite{BCS}, both for unconstrained metrics, and for metrics with constant scalar curvature.

It is well known that there exists a one parameter family of constant positive scalar curvature conformal metrics on $\mathbb R \times S^{n-1}$. In the literature, these metrics are usually referred to as Kottler--Schwarzschild--de Sitter metrics \cite{ChPollack} or Delaunay metrics \cite{MPU1} or even Fowler's metrics \cite{KMPS}. This family of metrics is parameterized by a parameter  which is called the ``neck size". Let us describe these briefly since they are at the heart of our result. We assume $n\ge 3$ and we set
\[
\epsilon_*: = \left( \frac{n-2}{n} \right)^{\frac{4}{n-2}} \, .
\]
For all $\epsilon \in (0, \epsilon_*]$, we define $v_\epsilon$ to be the unique solution of
\[
\partial_t^2 v_\epsilon - \left( \frac{n-2}{2} \right)^2 \, v_\epsilon  + \frac{n(n-2)}{4} \, v_\epsilon^{\frac{n+2}{n-2}} = 0 \, ,
\]
with $v_\epsilon = \epsilon$ and $\partial_t v_\epsilon (0) =0$. The parameter $\epsilon$ is called the \emph{neck size} of the associated
Delaunay metric
%
\bel{Delme}
g_\epsilon  = u_\epsilon^{\frac{4}{n-2}} \, dx^2 =  \, v_\epsilon^{\frac{4}{n-2}} \, ( dt^2 +  g_{S^{n-1}}) \,  ,
\ee
where
\[
u_\epsilon (x) : = |x|^{\frac{2-n}{2}}Ê\, v_\epsilon (-\log |x|) \, ,
\]
and where $dx^2$ denotes the Euclidean metric in $\mathbb R^n$ and $ g_{S^{n-1}}$ denotes the canonical metric on $S^{n-1}$. The metric $g_\epsilon$ has constant scalar curvature equal to
$R \equiv  n \, (n-1)$. Observe that, without loss of generality, we can normalize the metrics we are interested in to have (constant) scalar curvature equal to $n\, (n-1)$.

Building on the results in~\cite{Byde,ChDelay,ChPollack}, and using a simple perturbation argument, we prove the:
\begin{Theorem}
\label{Tmain}
Let $(M,g)$ be a smooth $n$-dimensional Riemannian manifold with constant positive scalar curvature $R  \equiv n\, (n-1)$, $n\ge 3$. Let $p\in M$ and suppose that there exist no local static KIDs near $p$. Then for any $\rho >0$ there exists $\epsilon_0>0$ such that for any $\epsilon\in (0,\epsilon_0]$ there exists a smooth metric with constant positive scalar curvature $R \equiv n \, (n-1)$ which coincides with $g$ in $M \setminus B_p(\rho)$ and which coincides with a Delaunay metric $g_\epsilon$ with neck size parameter $\epsilon$ in a punctured ball centered at $p$.
\end{Theorem}

As already pointed out, a generic metric will satisfy the ``no local static KIDs" property  at every point. Since both our construction and that of \cite{ChPollack} are purely local, the construction can be repeated at any chosen finite, or countably infinite, collection of points $p_i\in M$ to produce complete constant positive scalar curvature metrics with a countable number of exactly Delaunay ends.

Remark: It would be of interest to generalise this construction to general, not necessarily time-symmetric, general relativistic initial data sets $(M,g,K)$.

\subsection{Proof of Theorem~\ref{Tmain}:}
\label{sDlcf}

We choose $p \in M$ and $\rho >0$. There exists  $\bar \rho  \leq \rho$ such that for all $\delta < \bar \rho$  the linear operator
\[
L: = \Delta_{g} + n \, ,
\]
is injective on $\mathring {\mathcal C}^{2} (\bar B_p(\delta))$, the space of $\mathcal C^2$-functions which are defined on $\bar B_p(\delta)$ and which vanish on the boundary of this domain. (Observe that our Laplacian is the sum of second derivatives.)  Since we have assumed that there is no local static KIDs near $p$,  there exists $0 < \eta < \delta < \bar \rho$ such that there are no static KIDs on  the annulus
\[
A_p(\eta , \delta)  = \bar B_p (\delta ) \setminus B_p (\eta ) \, .
\]
Therefore, we conclude that we can choose  $ 0 < \eta < \delta \leq \rho$ such that there are no static KIDs on the annulus $A_p(\eta , \delta)$ and the operator $L$ is injective on $\mathring {\mathcal C}^{2} (\bar B_p (\delta))$.

In $\bar B_p(\delta)$, we now deform the metric $g$ to a family of metrics which are conformally flat in a small neighborhood $p$ and  which still have constant scalar curvature equal to $R = n \, (n-1)$ in $\bar B_p(\delta)$. The new metrics will match continuously the original ones along the boundary $\partial B_p(\delta )$, but in general their derivatives will not agree there.

\begin{Lemma}
\label{L1}
There exists $0 < r_0  \leq \eta $ and a family of constant scalar curvature metrics $g^r$, with $r \in (0,  r_0]$,  which are defined in $\bar B_p (\delta)$, have constant scalar curvature $R \equiv n \, (n-1)$, are conformally flat in $\bar B_p(r)$, the ball of radius $r$ centered at  $p$  (radius computed with respect to the metric $g$) and for which
\bel{redefL}
L_r:=\Delta_{g^r} + n \, ,
\ee
is injective on $\mathring {\mathcal C}^{2} (\bar B_p (\delta))$. Moreover
\begin{equation}
\|  g^r - g \|_{L^\infty(B_p (\delta))} + r \, \| \nabla_g (g^r - g) \|_{L^\infty(B_p (\delta))} \leq \, c \, r^{\gamma}
\label{est:1}
\;, \end{equation}
for any $\gamma  < 2$. Furthermore, for any $k\in \mathbb N$, the sequence of metrics $g^r$ converges to $g$ in $\mathcal C^k$-topology, on compacts of $\bar B_p(\delta) \setminus \{p\}$, as $r$ tends to zero.
\end{Lemma}

\noindent{\sc Proof:}
We agree that the geodesic balls, the gradient and norm are taken with respect to the metric $g$. We consider geodesic normal coordinates $x:=( x_1, \ldots, x_n )$ near $p$. In these coordinates, the metric $g$ can be expanded as
\[
g_{ij} =  \delta_i^j + {\mathcal O} (|x|^2)\, .
\]
We choose a cutoff function $\chi$ which is identically equal to $1$ in the unit ball of $\mathbb R^n$ and identically equal to $0$ outside the ball of radius $2$.  Given $ r \in (0 , \delta /2)$, we consider a metric $\bar g^r$ whose coefficients near $p$ are given by
\[
\bar g^{r}_{ij}: = \chi (\cdot /r ) \, \delta_{i}^j + (1  - \chi (\cdot / r ) ) \, g_{ij}
\;. \]

Observe that
\begin{equation}
\| \bar g^r -g \|_{L^\infty} +  r \,  \| \nabla_g ( \bar g^{r} -g) \|_{L^\infty} + r^2 \,  \| \nabla^2_g ( \bar g^{r} -g) \|_{L^\infty} \leq c \, r^{2}
\label{eq:1}
\;. \end{equation}
for some constant $c >0$ which does not depend on $r \leq \delta /2$.

Let us denote by $\bar R_r$ the scalar curvature of the metric $\bar g^{r}$ and recall  that $R =  n \, (n-1)$ is the scalar curvature of the metric $g$. We have $\bar R_r = R$ in $\bar B_p(\delta) \setminus B_p (2 r)$ and $\bar R_r =0$ in $\bar B_{p} (r)$. Finally, in the annulus $\bar B_{p} (2 r) \setminus B_p (r)$, we only have the estimate
\[
\| \bar R_{r} \|_{L^\infty} \leq c \, ,
\]
for some constant $c >0$ independent of $r \leq \delta /2$. This follows at once from the fact that the expression of the scalar curvature in terms of the coefficients of the metric involves the
coefficients of the metric and their derivatives up to order $2$ which, thanks to (\ref{eq:1})   are bounded independently of $r \leq \delta /2$.

Now, we explain how to solve the equation
\begin{equation}
\Delta_{ \bar g^r}  \, u - \frac{n-2}{4(n-1)} \, \bar R_{r} \, u + \frac{n \, (n-2)}{4}   \, u^{\frac{n+2}{n-2}} =0
\label{eq:2}
\;, \end{equation}
for $r$ small enough. Once this equation is solved and assuming that the function $u >0$, the metric
\[
g^ r:  = u^{\frac{4}{n-2}} \, \bar g^r \, ,
\]
will be a constant scalar curvature metric equal to $R \equiv  n \, (n-1)$ defined in $\bar B_p (\delta)$.

We set $ u = 1+v$ and rewrite the above equation as
\[
L   v =  (\Delta_g - \Delta_{ \bar g^r})  \, v +  \frac{n-2}{4(n-1)} \, ( \bar R_r - R) \, (1+v )  - \frac{n(n-2)}{4}  \, \left( (1+v)^{\frac{n+2}{n-2}} - 1 - \frac{n+2}{n-2} \, v\right)
\;. \]
For short, we denote by $N_r (v)$ the right hand side of this equation and we implicitly assume that the function $v$ is small enough, say  $\| v\|_{L^\infty} \leq 1/2$, so that $1+ v >0$.

We fix $\ell > n/2$. Some comments are due about the choice of $\ell$. Observe that functions if $W^{2, \ell}$ are in $L^\infty$ provided $\ell > n/2$ while their derivatives are in $L^\infty$ provided $\ell  > n$. In order to estimate the nonlinearities in the equation it will be easy to work with bounded functions, this justifies the choice $\ell > n/2$. Now, in the statement of the result, we claim some $L^\infty$ bound both on the solution itself and its partial derivatives, to do so, it would be tempting to work directly with $\ell >n$. However, with this latter choice we will obtain the desired $L^\infty$ estimate for the partial derivatives of the solution, but we will not obtain a good estimate for the $L^\infty$ norm of the solution itself; this is where $\ell > n/2$ will be needed.
 
 Using the fact that $\bar R_r - R$ is supported in $B_p (2r)$ and is bounded in this set, it is easy to check that there exists a constant $c>0$ (which does not depend on $r \in (0, \delta/2)$) such that
\[
\| ( \bar R_r - R ) \, (1+v ) \|_{L^\ell} \leq  c  \,  (1+ \| v\|_{L^\infty} ) \, r^{n/\ell}
\;,
\]
and
\[
\| ( \bar R_r - R) \, (v -v' ) \|_{L^\ell} \leq  c \, r^{n/\ell} \, \| v -v' \|_{L^\infty}
\;.
\]
Similarly,
\[
\|  (1+v)^{\frac{n+2}{n-2}} - 1 - \frac{n+2}{n-2} \, v \|_{L^\ell} \leq   c \, \| v\|_{L^\infty}^2
\;,
\]
and
\[
\|  (1+v)^{\frac{n+2}{n-2}} -  (1+v' )^{\frac{n+2}{n-2}} - \frac{n+2}{n-2} \, (v-v')  \|_{L^\ell} \leq   c \, (\| v \|_{L^\infty} + \| v '\|_{L^\infty}) \, \| v-v'\|_{L^\infty}
\;,
\]
provided $\| v\|_{L^\infty} \leq 1/2$ and $\| v '\|_{L^\infty} \leq 1/2$.

Thanks to (\ref{eq:1}) we also check that
\[
\| (\Delta_{g} - \Delta_{\bar g^r} ) \, v \|_{L^\ell} \leq \, c \, r \, \| v \|_{W^{2,\ell}}
\;.
\]
with perhaps another constant $c >0$ which again does not depend on $r \in (0, \delta/2)$. Indeed, the difference between these two operator involves a second order differential operator whose coefficients can be estimated by the difference of the coefficients of the two metrics and a first order differential operator whose coefficients can be estimated by the gradient of the difference between the coefficients of the two metrics.

By hypothesis, the operator
\[
L: = \Delta_{g} + n \, ,
\]
is an isomorphism from $\mathring W^{2,\ell} (B_p (\delta))$ into $L^{\ell} (B_p (\delta))$ (Recall that $\mathring W^{2,\ell} (B_p(\delta))$ is the completion of the space of smooth functions on $B_p(\delta)$ which  vanish on $\partial B_p(\delta)$ with respect to the usual $ W^{2,\ell}$-Sobolev norm).

Collecting these we find that
\[
\| L^{-1} \, N_r (v) \|_{W^{2,\ell }} \leq c \left(   \, r^{n/\ell}  + (r  + r^{n/\ell} ) \,  \| v \|_{L^\infty} +  \| v \|_{L^\infty}^2 \right)
\;,
\]
and
\[
\| L^{-1} \, (N_r (v)  - N_r (v'))\|_{W^{2,\ell }} \leq c \, \left(   r + r^{n/\ell} +  \| v\|_{L^\infty} + \| v' \|_{L^\infty} \right) \, \| v -v' \|_{L^\infty}
\;.
\]
Using the embedding $W^{2,\ell}(B_p (\delta)) \longrightarrow L^\infty (B_p (\delta))$, we conclude, from the fixed point theorem for contraction mappings, that the nonlinear operator
\[
v \mapsto  L^{-1} \, N_r (v)
 \]
has a (unique) fixed point  $v_r$ in the ball of radius $ 2 \, c \, r^{n/\ell} $ in $W^{2,\ell } (B_p (\delta))$, provided $r  >0$ is small enough, say $r \in (0, r_0]$.  This completes the proof of existence of a positive solution of (\ref{eq:2}). 
 The fact that $v$ (and hence $u$) is smooth follows from classical elliptic regularity.

The metric which appears in the statement of the result is
\[
g^r: = (1+ v_r)^{\frac{4}{n-2}} \, \bar g^r \, .
\]
The first estimate in (\ref{est:1}) follows from the construction itself and  the embedding  $W^{2,\ell} (B_p (\delta)) \longrightarrow L^\infty (B_p (\delta))$ when $\ell > n/2$,  while the second estimate in (\ref{est:1})  follows from the construction and the embedding  $W^{2,\ell} (B_p (\delta)) \longrightarrow W^{1, \infty} (B_p (\delta))$ when $\ell >n$.

Finally, reducing $r_0$ if necessary, we claim that the operator
\[
\Delta_{g^r} + n \, ,
\]
is injective for all $r \in (0, r_0]$. To do so, simply use (\ref{eq:1}) which implies that the coefficients of $g^r -g$ and their derivatives are  bounded in  $L^\infty$-norm  by a constant times $r^{\gamma-1}$. This implies that
\[
\| (\Delta_{g} - \Delta_{g^r} ) \, v \|_{L^\ell} \leq \, c \, r^{\gamma -1} \, \| v \|_{W^{2,\ell}}
\;.
\]
Finally,  the claim then follows from a simple perturbation argument since $L$ is an isomorphism between $\mathring W^{2,\ell} (B_p(\delta))$ and $L^{\ell} (B_p(\delta))$ for any $\ell >1$.

The fact that, for any $k\in \mathbb N$, the sequence of metrics $g^r$ converges to $g$ in $\mathcal C^k$-topology, on compacts of $\bar B_p(\delta) \setminus \{p\}$, as $r$ tends to zero follows from elliptic regularity theory since, for all $r \in (0, \bar r]$, the function $v_r$ is a solution of
\[
\Delta_{g}  \, v_r + n \, v_r + \frac{n \, (n-2)}{4}   \, \left(  (1+v_r)^{\frac{n+2}{n-2}}  - 1 - \frac{n+2}{n-2} \, v_r\right) =0
\]
in $\bar B_p(\delta ) \setminus B_p (\bar r)$  which vanishes on $\partial B_p (\delta)$ and is bounded by a constant times $r^\gamma$ in $L^\infty$-norm. \qed

\medskip

The following result is essentially borrowed from \cite{Byde}:

\begin{Theorem}
\label{TgenByde}
Let  $(M_0, g_0)$ be a compact Riemannian manifold with smooth boundary $\partial M_0$. Assume that the scalar curvature is constant equal to $R = n \, (n-1)$, and that the operator
\[
\Delta_{g_0} + n \, ,
\]
acting on functions $\mathring W^{2, \ell} (M_0)$ is injective. Further assume that the metric $g_0$ is locally conformally flat in a neighborhood of a point $p \in M_0$. Then, there exists $\epsilon_0 >0$   and for all $\epsilon \in (0,  \epsilon_0]$ there exists a complete constant scalar curvature metric defined in $M_0 \setminus \{ p \}$ of the form
\[
\tilde g_\epsilon  =e^{\phi_\epsilon} \, g_0\;,
\]
where $\phi_\epsilon =0$ on $\partial M_0$. Furthermore, $\phi_\epsilon$ converges to $0$ on compacts subsets of $M_0 \setminus \{p \}$ in any $
\mathcal C^k$-topology, for $k \in \mathbb N$.

\medskip

Finally, in the neighborhood of $p$ where $g_0$ is conformally flat and where we can use coordinates $x \in \mathbb R^n$ with $x=0$ at $p$, the metric $\tilde g_\epsilon$ can be written  as
\[
\tilde  g_\epsilon  = \tilde u_\epsilon^{\frac{4}{n-2}} \,  dx^2 \, ,
\]
where the function $u_\epsilon$ satisfies
\[
\tilde u_\epsilon (x) =  |x|^{\frac{2-n}{2}} \, \, v_\epsilon (-Ê\log |x| + t_\epsilon ) \, (1+ \mathcal O (|x|)) \, ,
\]
for some $t_\epsilon \in \mathbb R$.
\end{Theorem}

\begin{Remark}
In addition, reducing $\epsilon_0$ if   this is necessary, the solutions constructed are ``unmarked nondegenerate" (see section 6 of \cite{Byde} or Proposition
10 in \cite{MPa} for a proof). We refer to \cite{MPU1} for the definition of unmarked nondegeneracy. In particular this implies that one can use $p \in M_0$ and $\epsilon$ as coordinates on the unmarked moduli space.
\end{Remark}

\proof  Existence follows from Theorem 1.1 in \cite{Byde}. The fact that the construction is possible for any small value of the parameter $\epsilon$ is not explicit in the  statement of Theorem 1.1 in  \cite{Byde} but it is implicit in the proof (see bottom of page 1184). The expansion of the metric close to $p$  follows directly from the construction in \cite{Byde} but also follows from general results such as \cite{CGS} and \cite{KMPS}.
\qed

\medskip

Finally, we recall the result of  \cite[Theorem~5.9]{ChDelay}: if there are no static KIDs on the annulus $A_p(\eta , \delta)$, then there exists $\zeta >0$ such that if $\tilde g$ is a constant scalar curvature metric metric on the annulus  $\bar B_p(\delta) \setminus B_p(\eta/2)$ satisfying
\[
\| g- \tilde g\|_{\mathcal C^1 (A_p(\eta , \delta)) } \leq \zeta
\]
(recall that $A_p(\eta , \delta)=\bar B_p(\delta) \setminus B_p(\eta)$), then there exists  a constant scalar curvature metric $\hat g$  on $M$ which coincides with $g$ in $M\setminus B_p(\delta)$, and which equals $\tilde g$ in $\bar B_p(\eta) \setminus B_p(\eta/2)$.

We are now in a position to  prove Theorem~\ref{Tmain}.   We first apply Lemma~\ref{L1} and fix  $r \in (0, r_0]$ such that
\[
\| g - g^r \|_{\mathcal C^1 (A)} \leq \zeta /2 \, .
\]
At this stage $r$ is fixed and we next apply Theorem~\ref{TgenByde} with $M_0 = \bar B_p(\delta)$ and  $g_0 = g^r$ to obtain a family of asymptotically Delaunay metrics $\tilde g^{ r,\epsilon}$, with neck size $\epsilon\in (0,\tilde \epsilon_0]$.  Now, $g^{r, \epsilon}$ converges  to $g^r$ in $\mathcal C^k (A)$  as $\epsilon$ tends to zero. Therefore, reducing $\epsilon_0$ if necessary, we can assume that
\[
\| g^{r, \epsilon} - g \|_{\mathcal C^1 (A)} \leq \zeta \, .
\]
and \cite[Theorem~5.9]{ChDelay} applies to produce a constant scalar curvature metric $\mathring  g^{ r,\epsilon}$ which coincides with $g$ away from $B_p(\delta)$, and coincides with $g^{ r ,\epsilon}$ on $B_p (\eta)$. Observe that this metric is asymptotic to a Delaunay metric as explained in Theorem~\ref{TgenByde}. Finally,  applying Theorem 3.1 of \cite{ChPollack}, the  metric $\mathring g^{r ,\epsilon}$ can be deformed to a constant scalar curvature metric  which is \emph{exactly Delaunay} in the asymptotically Delaunay region. This completes the proof of our result.
\qed


\begin{thebibliography}{90}

\bibitem{BCS} R.~Beig, P.T. Chru\'{s}ciel, and R.~Schoen, \emph{{KIDs} are non-generic},
Ann.\ H.\ Poincar\'e \textbf{6} (2005), 155--194, arXiv:gr-qc/0403042.


\bibitem{Byde}
A.~Byde, \emph{Gluing theorems for constant scalar curvature manifolds},
 Indiana Univ.\ Math.\ Jour. \textbf{52} (2003), 1147--1199.

\bibitem{CGS}
L.A. Caffarelli, B.~Gidas, and J.~Spruck, \emph{Asymptotic symmetry and local
 behavior of semilinear elliptic equations with critical {S}obolev growth},
 Commun.\ Pure\ Appl.\ Math. \textbf{42} (1989), 271--297.

\bibitem{ChDelay}
P.T. Chru\'{s}ciel and E.~Delay, \emph{On mapping properties of the general
 relativistic constraints operator in weighted function spaces, with
 applications}, M\'em.\ Soc.\ Math.\ de France. \textbf{94} (2003), vi+103,
 arXiv:gr-qc/0301073v2.

\bibitem{ChPollack}
P.T. Chru\'{s}ciel and D.~Pollack, \emph{{Singular Yamabe metrics and initial
 data with \emph{exactly} Kottler--Schwarzschild--de Sitter ends}}, Ann.\
 H.~Poincar\'e (2007), in press, arXiv:0710.3365 [gr-qc].


\bibitem{KMPS}
N.~Korevaar, R.~Mazzeo, F.~Pacard, and R.~Schoen, \emph{Refined asymptotics for
 constant scalar curvature metrics with isolated singularities}, Invent.\
 Math. \textbf{135} (1999), 233--272.

\bibitem{MPa}
R.~Mazzeo and F.~Pacard, \emph{Constant scalar curvature metrics with isolated
 singularities}, Duke Math.\ J. \textbf{99} (1999), 353--418.

\bibitem{MPU1}
R.~Mazzeo, D.~Pollack, and K.~Uhlenbeck, \emph{Moduli spaces of singular
 {Y}amabe metrics}, Jour.\ Amer.\ Math.\ Soc. \textbf{9} (1996), 303--344.


\bibitem{Riess:2006fw}
A.G.~Riess \emph{et al.}, \emph{New {Hubble Space Telescope} discoveries of
 type {Ia Supernovae at $z > 1$: N}arrowing constraints on the early behavior
 of dark energy},  (2006), arXiv:astro-ph/0611572.

\bibitem{WoodVasey:2007jb}
W.M. Wood-Vasey et~al., \emph{Observational constraints on the nature of the
 dark energy: First cosmological results from the essence supernova survey},
 (2007), arXiv:astro-ph/0701041.

\end{thebibliography}
\end{document}